Review

# NSUN2 as a potential prognostic as well as therapeutic target in cancer by regulating m⁵C modification

By Zifang He [1,] Longtao Yang [2] and Peiyao Ma[3]

1 China Medical University, Shenyang, Liaoning Province
2 China Medical University, Shenyang, Liaoning Province
3 China Medical University, Shenyang, Liaoning Province

Correspondence should be addressed to: Shenbei Campus, China Medical University, No. 77, Puhe Road, Shenbei New District, Shenyang, Liaoning Province, China (e-mail: 20222218@cmu.edu.cn).

**Abstract:** m⁵C modification is a type of RNA methylation modification, and its major methyltransferase, NSUN2, catalyzes m⁵C modification. NSUN2 is overexpressed in a variety of cancers, and it affects the metabolism of RNA from target genes by affecting the level of m⁵C modification in cancer cells, which in turn promotes the development of cancers and is associated with poor prognosis. This review summarizes the mechanisms by which NSUN2 and m⁵C play a pro-cancer role in various cancers, and the relationship between NSUN2 and the prognosis of various cancers, with the aim of identifying NSUN2 as a prognostic indicator and a target for future cancer therapy, and to provide a clearer therapeutic idea and direction for the future treatment of cancer.

Non-standard abbreviation list: RB: Retinoblastoma; USUN: the NOP2/Sun-domain family; DNMT: the DNA methyltransferase family; TCGA: the Cancer Genome Atlas database; PFS: progression-free survival; DFS: disease-free survival; OS: overall survival; GC: Gastric cancer; EdU:5-ethynyl-2'-deoxyuridine; CRC: Colorectal cancer; NSCLC: Non-cancerous; IHC:

immunohistochemistry; BLCA: Bladder cancer; UCB: Urothelial bladder cancer; SYSUC: the Sun Yat-sen University Cancer Center databases; PCa: Prostate cancer; PRAD: Prostate adenocarcinoma; EC: Endometrial cancer; CC: Cervical cancer; KRT13: keratin 13;

## 1. Introduction

As cancer research continues to deepen, an increasing number of biomarkers associated with tumorigenesis and the relationship between RNA modifications and cancer are being discovered and confirmed. RNA modifications play a key role in many biological processes. As of 2017, 163 post-transcriptional RNA modifications have been discovered [1], including A to I Editing(A-I), N4-acetylcytidine(ac4C),5-methylcytosine (m5C), N6 methyladenosine (m6A), and pseudouridine (ψ) (Fig.1), and this list continues to grow [1], with more RNA modifications to be discovered in the future. Currently, RNA post-transcriptional modifications in tumors are attracting increasing attention [2,3,4,5].

m5C modification is widely found in DNA and RNA [6] and can be involved in RNA export, ribosome assembly [16], maintenance of RNA stability, and regulation of RNA translation efficiency [17,18]. Aberrant m5C modifications have been observed in many tumors. For example, in retinoblastoma (RB) cells, the modification of mRNA by m5C is higher than that in normal tissues or cells [19,20] and differences in the distribution of m5C have also been found in hepatocellular carcinoma and paraneoplastic tissue [21,22]. The relationship between m5C and tumors has attracted increasing attention [23].

The entire m5C modification process is catalyzed by a methyltransferase ("writer"), which transfers methyl groups to the cytosine bases of the RNA [24], and then is followed by demethylases ("wipers") and binding proteins ("readers ") are dynamically regulated [25]. NSUN2 is an m5C -modified methyltransferase [27] belonging to the NOP2/Sun-domain (NSUN) family. Currently, all known mammalian m5C RNA methyltransferases belong to either the DNA methyltransferase (DNMT) family or the NSUN family (NSUN1-NSUN7), and members of both families work together to add m5C to RNA cytosines [27] (Fig.1). Of all the m5C-modified methyltransferases, NSUN2 has been extensively studied [28] and has been reported to be overexpressed in a variety of cancer types, including gastric, bladder, gallbladder, and breast cancers [31,32]. We also performed a bioinformatics analysis of NSUN2 expression in cancer based on The Cancer

Genome Atlas (TCGA) database, which also verified this finding. Overall, NSUN2 promoted the formation of m5C during mRNA translation and catalyzed the methyl addition process of RNA, which in turn promoted the expression of downstream target genes, which in turn promoted proliferation, migration, and invasion, leading to further development of cancer (Fig.3).

We also found that NSUN2 is not only upregulated and expressed in cancer but is also associated with cancer prognosis. By analyzing the clinical data of progression-free survival (PFS), disease-free survival (DFS), overall survival (OS), and clinicopathological manifestations of various cancers such as gastric, colorectal, bladder, and prostate cancers in the TCGA and GEO databases, we were able to find that the high expression of NSUN2 is usually accompanied by poor prognosis of cancer patients and the higher the level of expression, the worse the prognosis, and in some cancers, the higher the level of expression, the worse the prognosis. The higher the level of NSUN2 expression, the worse the prognosis; in some cancers, the expression of NSUN2 was also observed to be related to the tumor size and TNM stage of the cancer [37], such as NSCLC [54]. This suggests that it is possible to combine NSUN2 with clinical data and designate NSUN2 as a clinical prognostic indicator, reflecting the degree of cancer progression through the expression level of NSUN2 as a means of subsequent treatment. At the same time, it can also be reasonably hypothesized that if the expression of NSUN2 is reduced, the prognosis of patients can be improved to a large extent, and NSUN2 will also become a therapeutic target for cancer, which remains to be investigated in the future.

In this review, we summarize the developmental patterns of NSUN2 and m5C in a variety of cancers and the relationship between NSUN2 expression and clinical prognosis, including gastric, colorectal, lung, bladder, prostate, endometrial, cervical, and ovarian cancers. By studying the relationship between NSUN2 and m5C and clinical prognosis in these cancers, we can further deepen our understanding and knowledge of NSUN2 and m5C and provide more ideas and directions for future cancer treatment.

## 2. NSUN2 exerts pro-oncogenic effects and decreases clinical prognosis in an m5C -dependent manner in a variety of cancers

### 2.1. Cancers of the digestive system.

*2.1.1. Gastric cancer.* Gastric cancer (GC) is among the most common malignancies worldwide, occupying the fourth position in mortality and fifth in incidence [3]. It constitutes the third primary cause of cancer-related deaths [3], with China accounting for more than 40% of new cases and fatalities [36]. Despite the development of numerous therapeutic strategies and improvements in the overall survival rate for GC, the prognosis for patients with metastatic, recurrent, or advanced-stage disease remains poor. Therefore, there is a pressing need to decipher the mechanisms underlying GC progression to develop novel therapeutic approaches.

Researchers have found that the expression levels of several enzymes involved in depositing (writers) and detecting (readers) m5C modifications are higher in GC samples than those in normal tissues, with NSUN2 showing the most prominent expression. This observation implies that NSUN2 might act as an oncogene in GC and is correlated with poor prognosis in patients with GC [37]. Furthermore, gene sequencing data indicated that m5C regulatory factors are frequently overexpressed in GC samples, and the overall level of their expression is significantly associated with patient survival outcomes. To explore the oncogenic function of NSUN2 in GC cells, we silenced NSUN2 in these cells and subsequently performed CCK-8 and 5-ethynyl-2′-deoxyuridine (EdU) incorporation assays. The results showed that decreasing NSUN2 expression resulted in a reduced proliferation rate of the cell lines, in contrast to the control group, in which NSUN2 overexpression markedly enhanced cell proliferation. To evaluate the prolonged effect of NSUN2 on GC cell proliferation, a colony formation assay was performed, which suggested that NSUN2 promoted the proliferation of GC cells [27]. This conclusion was supported by additional studies [38].

To further investigate whether the oncogenic function of NSUN2 relies on its m5C methyltransferase activity and to assess its relationship with m5C, researchers constructed two plasmids, one harboring the wild-type NSUN2 gene and the other carrying a mutant NSUN2 gene with point mutations at two critical sites that abolish its m5C enzymatic function, creating an enzyme-inactive mutant [39]. By introducing both plasmids into NSUN2-knockdown cell lines, researchers observed that cells transfected with the mutant NSUN2 plasmid demonstrated markedly decreased proliferation and metastatic potential. These results suggest that NSUN2 facilitates cell proliferation and metastasis, and consequently, gastric cancer progression predominantly occurs through an m5C-mediated mechanism [27]. Meanwhile, RNA double sequencing analysis was performed, which revealed 3,260 m5C peaks in wild-type cells versus 712 m5C peaks in NSUN2-deficient cells. This discovery further supports the idea that NSUN2 is the primary enzyme responsible for m5C modification of eukaryotic mRNAs [27]. Similar results have been reported in other studies, where NSUN2 overexpression was observed in gastric cancers (GCs), and it was verified that NSUN2 enhances GC cell proliferation by inhibiting the downstream target gene p57 (Kip2) in an m5C-dependent manner [38].

Some investigators also performed Kaplan-Meier survival analysis, which showed that the overall survival (OS) of GC patients with high NSUN2 expression was lower than that of GC patients with low NSUN2 expression [27]. In conclusion, all of the above studies have demonstrated that the RNA methyltransferase NSUN2 is highly expressed in GC and that its downregulation can inhibit GC cell proliferation and metastasis in vitro, whereas NSUN2 overexpression could promote the proliferation and metastasis of GC cells. It was also found that NSUN2 knockdown significantly reduced m5C levels in GC cells, further demonstrating that m5C modification is mainly catalyzed by NSUN2 [27] and that elevated expression of NSUN2 is associated with poor prognosis in GC patients, with poor prognosis in patients with high NSUN2 expression.

*2.1.2 Colorectal cancer.* Colorectal cancer (CRC) is the third most common malignancy globally [41], with over 1.9 million new cases diagnosed each year [42]. The malignancy is highly aggressive, and the 5-year survival rate for patients with metastatic CRC is merely 13.1% [16]. Hence, gaining thorough insight into the mechanisms underlying CRC progression and metastasis is crucial for the development of targeted therapies.

In TCGA database, a differential expression pattern of NSUN2 and m5C was observed, with notably higher protein levels in CRC tissues than in adjacent normal tissues [43]. Comparable observations and phenomena have been documented in previous studies [45, 46, 47, 48, 49]. Additionally, immunofluorescence assessment of CRC cells with NSUN2 knockdown and overexpression demonstrated a notable positive association between m5C fluorescence intensity and NSUN2 expression. This result implies that NSUN2 can modulate the degree of m5C modification in CRC cells [44].

To assess the oncogenic potential of NSUN2 in colorectal cancer (CRC), two CRC cell lines with NSUN2 knockdown were generated. Functionally, knockdown of NSUN2 markedly suppressed the proliferation and invasiveness of CRC cells. In contrast, NSUN2 overexpression exerted the opposite effect [43]. Moreover, re-expression of NSUN2 rescued CRC cell growth. In line with this observation, colony formation assays showed that NSUN2 enhanced colony formation, and the number of migrating cells decreased significantly upon NSUN2 knockdown. This indicated that elevated NSUN2 levels facilitated cell migration. Collectively, these findings further substantiate the tumor-promoting role of NSUN2 [44].

Building on these findings, four mouse models were established to examine the tumorigenic impact of NSUN2 in vivo. As anticipated, nude mice inoculated with CRC cells lacking NSUN2 exhibited a marked decrease in subcutaneous tumor volume and weight, along with a reduced number and size of lung and liver metastases compared to control mice. Restoration of NSUN2 expression reverses these outcomes [44]. Taken together, these results indicate that NSUN2 stimulates CRC growth both in vivo and in vitro, and plays a pivotal role in CRC tumor progression and metastasis [43].

Correlational studies examining NSUN2 in CRC clinicopathology revealed a significant association between NSUN2 expression levels and both tumor size and TNM stage in CRC. Furthermore, Kaplan-Meier survival analysis indicated that CRC patients with high NSUN2 expression had reduced overall survival (OS) compared to those with low NSUN2 expression [44]. At the same time, we also used TCGA database to analyze the survival of patients with CRC, and the results showed the same trend as in previous studies; patients with high expression of NSUN2 had lower PFS (Fig. 4), although the results were not statistically significant. We believe that, when the number of samples is sufficiently large, it will surely make this trend is more obvious. In other studies, there were also results showing that patients with high NSUN2 expression had lower DFS, suggesting that high NSUN2 expression is associated with recurrence, and at the same time, this researcher also performed Multivariate Cox regression analysis, and the results showed that elevated NSUN2 expression was an independent prognostic indicator for OS [73]. In summary, these findings imply that NSUN2 contributes to cancer progression in CRC and may serve as a prognostic marker for patients with CRC in clinical settings.

## 2.2 Cancers of the respiratory system

*2.2.1 Lung cancer.* Lung cancer, a malignant tumor that occurs in the bronchial mucosa or lung glands, is highly prevalent and on the rise worldwide [3]. Histopathological analysis indicates that approximately 85% of lung cancer cases are classified as non-cancerous (NSCLC), which includes two primary subtypes: adenocarcinoma of the lung (LUAD) and squamous cell carcinoma (LUSC) [51]. Currently, there is a crucial demand for the identification of new prognostic markers and therapeutic options to improve the prognosis of patients with early-stage lung cancer.

NSCLC is a highly diverse, aggressive, and progressive disease with limited treatment options and low survival rates [52,53]. RNA-seq data obtained from TCGA and GEO databases showed increased NSUN2 expression in NSCLC tissues relative to normal tissues. Subsequently, the research team examined the transcriptomic and survival data obtained from these databases. Their analysis revealed that NSCLC patients with higher NSUN2 expression had worse prognostic outcomes, notably in stages I and IV. Furthermore, the researchers employed immunohistochemistry (IHC) and western blot (immunoblotting) methods to evaluate NSUN2 expression in normal and NSCLC tumor tissues sourced from hospitals. These findings indicated that NSUN2 expression was upregulated in NSCLC tissues, with higher levels tightly linked to tumor stage and size. Meanwhile, other researchers analyzed the correlation between NSUN2 and clinicopathological parameters based on NSUN2 expression and clinical data, and the results also showed that high levels of NSUN2 expression were positively correlated with NSCLC tumor grade and size [54].

To investigate the biological significance of NSUN2 in non-small cell lung cancer, researchers conducted in vitro functional experiments. Using lentiviral transfection, researchers inhibited the expression of NSUN2 in cells. The CCK8 assay showed that inhibition of NSUN2 decreased the survival of lung cancer cells compared to that of control cells. EdU and colony formation assays were performed to investigate the effect of NSUN2 on cell proliferation in NSCLC. Notably, when NSUN2 was knocked down, a decrease in EdU double-labelled proliferating cells and in the number of colonies was observed, suggesting that the lack of NSUN2 significantly impaired the proliferative potential of NSCLC cells, which is consistent with the findings in gastric cancer. Given that cancer-related deaths are often attributed to the metastasis and dissemination of cancer cells, researchers have assessed the impact of NSUN2 on NSCLC cell migration and invasion capabilities. The wound healing assay results demonstrated that suppressing NSUN2 significantly decreased cellular motility when compared to the control group and also hindered the cells' migratory ability. Taken together, these results imply that the knockdown of NSUN2 hinders the advancement of non-small cell lung cancer in vitro, implying an oncogenic function for NSUN2. Conversely, the opposite outcome was observed following the establishment of NSCLC cell lines with elevated NSUN2 expression via lentiviral transfection. The CCK-8 assay indicated heightened cell viability, the EdU proliferation assay revealed an increased ratio of EdU-positive cells, and the colony count in the NSUN2-overexpressing group was markedly higher than that in the control. Additionally, the cell cycle duration was reduced, and the invasive capacity was significantly augmented. These findings imply that ectopic expression of NSUN2 accelerates NSCLC progression by stimulating cell proliferation, migration, and invasion [54].

Researchers examined 497 LUAD samples and 54 specimens of normal lung tissue from the TCGA database. Their findings indicated that the expression of multiple genes was notably upregulated in LUAD tissues relative to adjacent normal tissues, with NSUN2 being prominent. Furthermore, Kaplan-Meier survival analysis indicated that individuals with elevated NSUN2 expression experienced decreased overall survival (OS). Furthermore, NSUN2 levels are elevated in lung adenocarcinoma, accompanied by a significant increase in m5C levels [55]. This finding further strengthens the robust positive correlation between m5C and NSUN2 levels.

### 2.3 Cancers of the urinary system

*2.3.1 Bladder Cancer.* Bladder cancer (BLCA), a significant global health concern, is the 10th most common malignancy worldwide in terms of incidence [56], contributing to roughly 170,000 fatalities annually [57]. This highly aggressive genitourinary neoplasm often presents insidiously and is prone to misdiagnosis, with rising morbidity and mortality rates in recent years [58]. The pathogenesis of BLCA is intricate, involving multiple factors and steps in a complex pathological process, and the underlying mechanisms of disease progression are yet to be fully understood.

To investigate the expression profile of NSUN2 in BLCA, researchers examined both protein and RNA samples derived from BLCA and healthy tissues. The findings indicated that NSUN2 protein expression was markedly elevated in BLCA tissues compared to that in normal tissues, mirroring the pattern observed at the mRNA level. Furthermore, NSUN2 protein and mRNA levels were notably higher in BLCA cell lines than in normal uroepithelial cell lines [55]. Additionally, data obtained from TCGA corroborated the overexpression of NSUN2 in BLCA [55].

Analysis of survival data revealed a significant association between high NSUN2 expression and reduced overall survival [60]. We also analyzed the data in TCGA database and found that patients with high NSUN2 expression also had lower PFS and DFS (Fig.5 and 6), which is consistent with our previous conclusion that high NSUN2 expression is associated with a poor prognosis. Although the results of our bioinformation analysis were not statistically significant, we believe that an increase in the sample size would certainly lead to fuller confirmation. Examination of the TCGA database further demonstrated that NSUN2 was linked to increased or decreased tumor risk. Moreover, Spearman's correlation analysis indicated a notable negative correlation between NSUN2 protein levels and survival duration, suggesting that elevated NSUN2 expression correlates with shorter survival times [60].

Analysis of urothelial bladder cancer (UCB), a subset of bladder cancer, has shown that NSUN2 expression is frequently upregulated in UCB tumor tissues. This was determined by examining the expression patterns of m5C regulators in UCB, using data from both the Sun Yat-sen University Cancer Center (SYSUCC) and TCGA databases, which included RNA data from normal tissues and adjacent UCB tumor tissues. Additionally, by combining RNA sequences from UCB cells and tumor specimens, a positive correlation was observed between the m5C level of NSUN2 mRNA and NSUN2 RNA expression in the SYSUCC dataset [61]. These findings further implicate NSUN2 in cancer through regulation of m5C.

*2.3.2 Prostate cancer.* Prostate cancer (PCa), also known as prostate adenocarcinoma (PRAD), is a major health issue for males. Globally, it ranks as the second most common malignancy among males and the fifth primary reason for cancer-related deaths in males. Statistics from 2020 reveal around 1.4 million new instances and 375,000 fatalities [3,62]. In Western nations, it is the most common male cancer, and its occurrence is escalating rapidly in Asia.

In TCGA dataset, NSUN2 emerged as the most abundantly expressed m5C methyltransferase gene in prostate cancer (PCa). TCGA study findings showed that the levels of NSUN2 mRNA were significantly elevated in tumor specimens compared to those in normal tissues. Furthermore, compared to adjacent normal tissues, the expression of NSUN2 mRNA showed a marked increase in tumor tissues. Both qPCR and immunohistochemistry assessments of NSUN2 expression in PCa and normal tissues corroborated the increased NSUN2 levels in PCa [65]. Other studies have also reported high NSUN2 expression in PCa [66,67]. Furthermore, researchers observed that heightened NSUN2 expression in PCa patients was correlated with reduced survival. Patients with high NSUN2 expression have been shown to have lower PFS, DFS, and OS. Collectively, these findings indicate that patients with high NSUN2 expression have poor clinical prognosis. NSUN2 upregulation is a prevalent oncogenic feature in PCa patients [64].

To determine the potential of NSUN2 as a therapeutic target in prostate cancer (PCa), researchers have conducted a series of experimental studies. They established cell lines using lentiviral transfection technology to selectively target and silence NSUN2. Colony formation assays, CCK-8 assays, and transwell assays were then performed, and the results showed that NSUN2 played an important role in promoting PCa cell proliferation and invasion. Additionally, wound healing showed a significant increase in cell migration in the overexpression of NSUN2 group and a decrease in cell migration in the NSUN2 knockdown group. In vivo studies also showed a significant increase in tumor volume in the NSUN2 overexpression group and a decrease in tumor volume in the NSUN2 knockdown group. Collectively, these results suggest that NSUN2 affects the proliferation, invasion, and migrate in laboratory cultures and organisms [64]. Similar observations have been reported in other studies [65], further supporting the notion that NSUN2 is critical for PCa progression and may be a promising therapeutic approach.

### 2.4 Cancers of the female reproductive system

*2.4.1 Endometrial Cancer.* Endometrial cancer (EC), a gynecological malignancy that arises from the endometrium, is witnessing a gradual increase in its incidence globally [68]. Approximately 90% of EC cases are identified in the early stages, and the 5-year survival rate for these patients is relatively high [69]. Currently, patients with recurrent EC face a poor outlook due to the lack of efficacious adjuvant therapy [70]. Consequently, there is a pressing need to develop novel therapeutic strategies.

To explore the potential involvement of $m^5C$ modifications in the progression of endometrial cancer (EC), researchers examined the expression patterns of key $m^5C$ regulators within the TCGA-UCEC database. They discovered that NSUN2, an enzyme responsible for catalyzing $m^5C$ formation in mRNA, was markedly upregulated. Analysis of NSUN2 mRNA and protein expressions in TCGA database indicated significantly elevated levels in EC tissues relative to adjacent normal tissues [71]. To confirm these observations, RT-qPCR and western blotting were performed on surgically resected samples to quantify NSUN2 mRNA and protein levels. These results support the finding that NSUN2 is overexpressed in EC tissues compared to normal tissues, aligning with immunohistochemical data [72].

To deepen the understanding of NSUN2's role in the advancement of endometrial cancer (EC), three cell lines were generated via viral transfection, followed by purification of their mRNA. These findings revealed that reducing NSUN2 expression led to a decrease in $m^5C$ levels, whereas increasing NSUN2 expression increased $m^5C$ mRNA levels. Following this, a range of experiments were conducted to assess the impact of NSUN2 on the proliferation of EC cells, using CCK-8, colony-forming, and EdU incorporation assays. These findings indicate that knocking down NSUN2 significantly inhibited cell proliferation in the cell lines, while overexpressing NSUN2 enhanced cell growth. Subsequently, the research team generated wild-type and double-mutant NSUN2 overexpression constructs and introduced them into EC cells that had undergone stable NSUN2 knockdown. The mutant NSUN2 was unable to restore the proliferative capacity of EC cells to the level observed with the wild-type NSUN2. Collectively, these findings indicate that NSUN2 enhances EC proliferation in an in vitro setting. Additionally, silencing of NSUN2 led to a significant decrease in $m^5C$ modification sites. further suggested that NSUN2 facilitates EC progression through the regulation of $m^5C$ modification [71].

During the investigation, it was observed that NSUN2 expression in endometrial cancer (EC) demonstrated a robust correlation with FIGO stage, histological grade, and pathological subtype, based on clinicopathological characteristics. Previous studies have shown that patients with high NSUN2 expression have lower OS [74], and to make this claim more convincing, we also conducted a survival analysis of EC clinical patients using the TCGA database, the results showed that patients with high NSUN2 expression also concurrently had lower PFS (Fig. 7). Furthermore, the study revealed that patients exhibiting elevated NSUN2 levels had poorer outcomes. Together, these results indicate that NSUN2 is often upregulated in EC, underscoring its potential as a prognostic indicator [72].

*2.4.2 Cervical cancer.* Cervical cancer (CC) is one of the most prevalent gynecological malignancies worldwide, with annual figures showing over 600,000 new diagnoses and 340,000 fatalities globally by 2020 [3,30,34]. Despite the existence of effective prevention measures, including HPV vaccination and cervical cancer screening, CC remains a substantial health issue, especially in low- and middle-income nations. Notably, a considerable proportion of cervical cancer cases are detected at advanced stages [26]. Consequently, CC poses a significant medical challenge, underscoring the importance of identifying novel potential targets for more efficacious treatment and management approaches. Elucidating the underlying molecular pathways and identifying key factors involved in cervical cancer progression could pave the way for targeted therapies, enhancing patient outcomes, and alleviating the overall disease burden.

Studies have revealed that increased NSUN2 levels significantly enhance the proliferation, migration, and invasive capacity of cervical cancer (CC) cells. In contrast, suppressing NSUN2 markedly disrupted CC cell proliferation, migration, and invasive potential. To delve deeper into the oncogenic function of NSUN2 in vivo, researchers have performed subcutaneous tumorigenesis experiments. These experiments were designed to evaluate the effect of NSUN2 expression on tumor growth and progression in live organisms, offering insightful data on NSUN2's potential as a therapeutic target for cervical cancer. When NSUN2 was knocked down, nude mice developed smaller and lighter xenograft tumors compared to the control groups. Immunohistochemical (IHC) analysis demonstrated reduced levels of Ki67, a proliferation marker, in tumors originating from NSUN2-knockdown cells. Moreover, the mRNA levels of NSUN2 were significantly elevated in cervical cancer tissues compared to those in normal tissues. These findings imply that NSUN2 is overexpressed in cervical cancer and plays a pivotal role in its oncogenesis [63].

LRRC8A, a volume-regulated anion channel protein crucial for cellular homeostasis, exhibits elevated expression in cervical cancer (CC). RNA BisSeq analysis has implicated m5C modifications in the regulation of LRRC8A expression [35]. To investigate this further, researchers inhibited NSUN2, an m5C methyltransferase, and noted a decrease in the m5C modification of LRRC8A mRNA. This finding suggests that NSUN2 modulates LRRC8A expression, its downstream gene, via m5C modifications [63]. Moreover, evidence shows that NSUN2 enhances CC cell migration and invasion by promoting m5C methylation of keratin 13 (KRT13) mRNA [79]. This supplementary data further underscores NSUN2's role as an m5C-mediated regulator, emphasizing its potential as a therapeutic target for modulating gene expression and halting cancer progression.

To explore the relationship between NSUN2 and clinical patient prognosis, some researchers conducted a survival analysis of CC clinical data in the GEO database, and the results showed that patients with high expression of NSUN2 had worse OS [75]. These results indicate that patients with elevated and high expression of NSUN2 in CC have poor prognosis, which provides another strong basis for the designation of NSUN2 as a clinical prognostic indicator.

### 3.Conclusion

This review summarizes the expression of NSUN2 in many different cancers, including gastric, colorectal, lung, bladder, prostate, endometrial, cervical and ovarian. Overall, we found that NSUN2 is highly expressed in various cancers. NSUN2 is associated with tumor growth, and its high expression promotes the production of tumor cell proteins and tumor growth, as well as enhances the proliferation, migration, and invasion of tumor cells. In clinical practice, it can also be found that high expression of NSUN2 is usually associated with poor prognosis of cancer patients, so it can be used as a prognostic indicator and therapeutic target of cancer. At the same time, to clarify the pro-cancer role of NSUN2 and the mechanism of cancer, it can also be a very good way to improve the prognosis of cancer patients as well as the survival of cancer patients, and to provide help for the treatment of cancer. This paper also summarizes the relationship between NSUN2 and m⁵C in cancer, as well as the mechanism by which they play a role in cancer. In general, NSUN2, as a kind of m⁵C RNA methyltransferase, is able to promote the formation of m⁵C modification, and usually acts in an m⁵C-dependent manner by adding methyl groups to the target genes, which affects the expression of the target genes, and then influences the development of the disease. In summary, although the functional role of NSUN2-catalyzed m⁵C methylation has not been fully described, and its specific mechanism remains to be investigated, current evidence has shown that NSUN2-mediated methylation has an important role and is an important research target for various types of cancer, NSUN2 represents a critical prognostic marker for cancer, and elucidating the mechanism of action of NSUN2 and m⁵C may provide a promising strategy for targeted cancer therapy.

## Acknowledgements

We thank Peiyao Ma for reviewing and providing feedback on the initial draft. Figures 1and 2 were created using Biorender.

## Conflict of interest

The authors declare no competing interests.

Figure Legend

Fig. 1. schematic representation of the structural location of RNA modifications on genes and the DNMT and NSUN families as methyltransferases promoting m5C formation, m5C is more aggregated in the 5'UTR and 3'UTR regions, and the DNMT and NSUN families can catalyze the process of adding methyl groups to genes, i.e., catalyze the formation of m⁵C. (Created in https://BioRender.com)

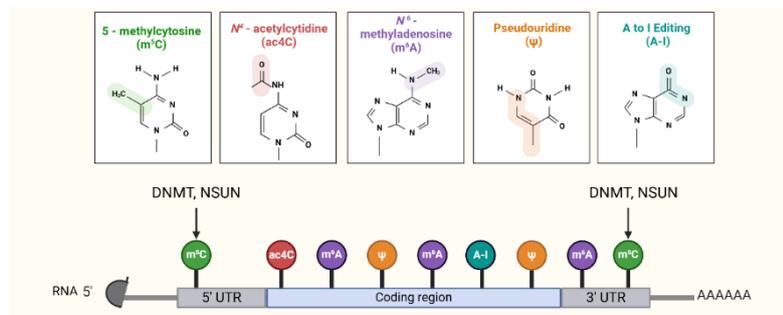

Fig. 2. The pan-cancer expression status of NSUN2 mRNA in TCGA database (normal vs. tumor) shows that NSUN2 is overexpressed in a variety of cancers compared to normal tissues.

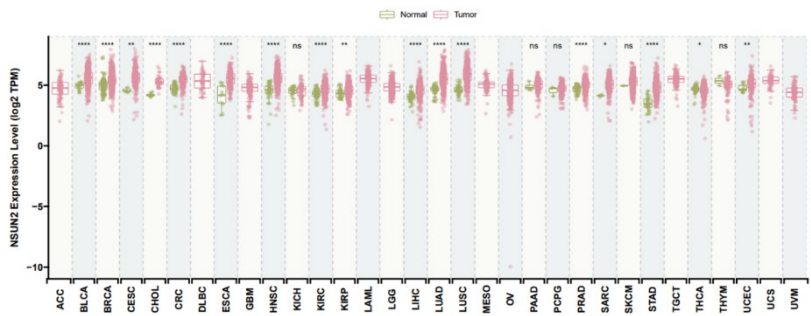

Fig. 3. Schematic representation of the mechanism of NSUN2 and m5C in gastric, colorectal, lung, bladder, prostate, endometrial, cervical, ovarian and osteosarcoma. NSUN2 catalyzes the formation of m5C methylation during mRNA translation and promotes the expression of corresponding proteins, which in turn drives the development of cancer. (Created in https://BioRender.com)

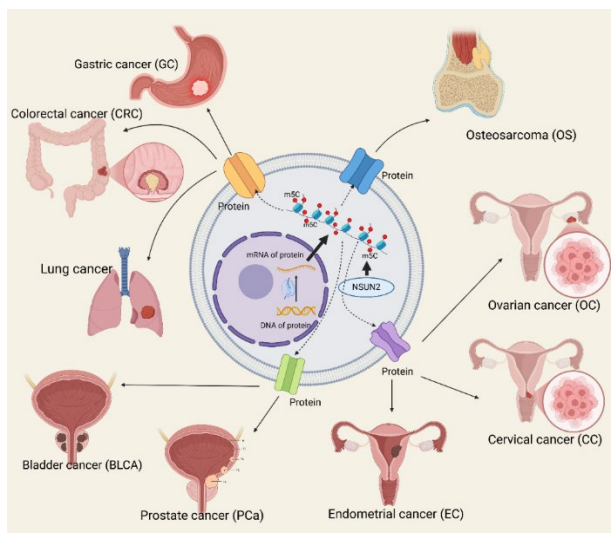

Fig. 4. K-M curves of the PFS of 189 patients from the TCGA cohort. The survival analysis is made by the data from TCGA-CRC. The patients were divided into high-NSUN2 and low-NSUN2.

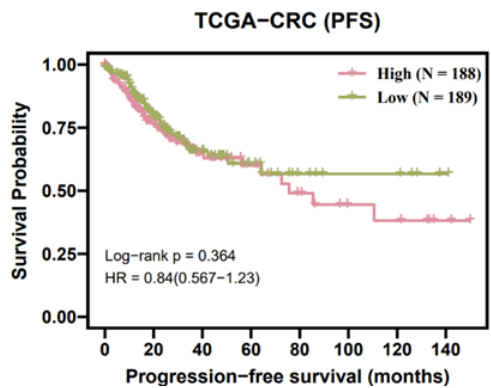

Fig. 5. K-M curves of the PFS of 203 patients from the TCGA cohort. The survival analysis is made by the data from TCGA-BLCA. The patients were divided into high-NSUN2 and low-NSUN2.

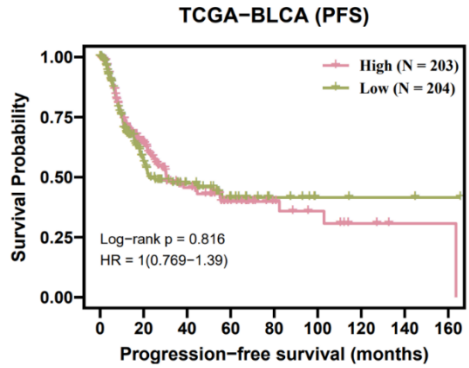

Fig. 6. K-M curves of the DFS of 203 patients from the TCGA cohort. The survival analysis is made by the data from TCGA-BLCA. The patients were divided into high-NSUN2 and low-NSUN2.

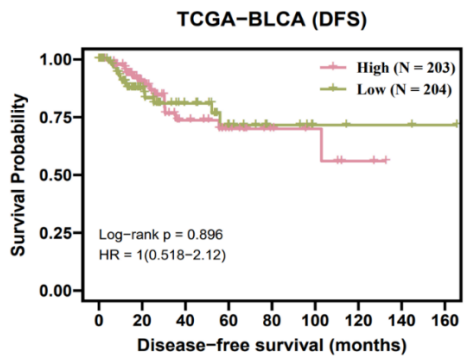

Fig. 7. K-M curves of the PFS of 277 patients from the TCGA cohort. The survival analysis is made by the data from TCGA-UCEC. The patients were divided into high-NSUN2 and low-NSUN2.

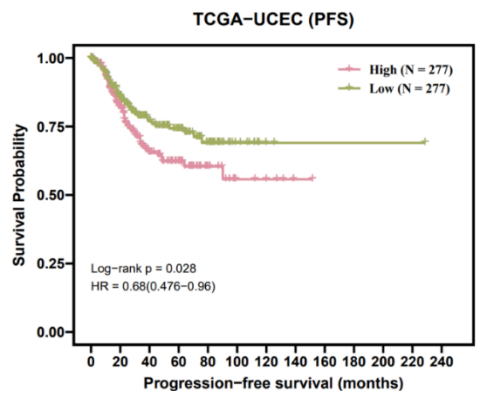